\documentclass[onecolumn, 10pt, a4paper]{article}
\usepackage[dvips]{graphicx}
\begin{document}
\title{\bf Liquid Xe scintillation calorimetry and Xe optical properties} 
%

\author{A.~Baldini$^a$, C.~Bemporad$^a$, F.~Cei$^a$, T.~Doke$^d$, \\
        M.~Grassi$^a$, T.~Haruyama$^b$, S.~Mihara$^c$, T.~Mori$^c$, \\
        D.~Nicol\`o$^a$, H.~Nishiguchi$^a$, W.~Ootani$^c$, K.~Ozone$^c$,\\ 
        A.~Papa$^a$, R.~Pazzi$^a$, R.~Sawada$^c$, F.~Sergiampietri$^a$, \\
        G.~Signorelli$^a$, S.~Suzuki$^d$, K.~Terasawa$^d$\\[1cm]
$^a$ {\it INFN Pisa, University and Scuola Normale Superiore di Pisa, Italy}\\ 
$^b$ {\it Institute of Particle and Nuclear Studies, KEK, Tsukuba, Japan }\\
$^c$ {\it ICEPP Tokyo and Tokyo University, Japan}\\
$^d$ {\it Waseda University, Tokyo, Japan}\\
}
\date{\emph{Revised August $28^{th}, 2003$}}

\maketitle
%
%
\begin{abstract}
The optical properties of LXe in the vacuum ultra violet (VUV), determining the performance of a scintillation calorimeter, are discussed in detail. The available data, measured in a wider spectral region from visible to UV light, and in a large range of Xe densities, from gas to liquid, are examined. It is shown that this information can be used for deriving the LXe optical properties in the VUV. A comparison is made with the few direct measurements in LXe for VUV light resulting from the LXe excitation by ionizing particles. A useful relation is obtained which connects the Rayleigh scattering length to the refractive index in LXe. 
\end{abstract}

\section{Introduction}
\label{sec:intro}
An 800-liter LXe (liquid xenon) scintillation $\gamma$-detector is being 
developed for the MEG experiment, which will search for the 
$\mu^{+} \to e^+ \gamma$ decay at the Paul Scherrer 
Institute~\cite{proposal}.

The MEG calorimeter is designed to achieve superior 
performances such as $\Delta E_{\gamma} = 4\%$ for the $\gamma$~energy,  
a position resolution $\Delta x_{\gamma} \approx 5$ mm, and $\Delta t_{e\gamma} = 150$~ps for timing, where $\Delta$ stands for the 
FWHM resolution.
The LXe calorimeter is only based on scintillation light without 
any attempt to measure ionization. The final detector will have an active
volume of 800-liter LXe and 800 PMTs.
Since no such large size LXe detector was ever produced, smaller prototypes were made to gain a know-how in LXe calorimetry. The first prototype had an active volume of 2.34 liter LXe and 32 PMTs \cite{mihara}. 
A second prototype has an active volume of 68.6 LXe and it is an important milestone in view of the final detector, since it allows to
gain a practical knowledge of the behaviour of such a device.  Above all,
the purification of LXe and  the long-term stability are essential for
ensuring a high light yield and a high performance of the detector. 
 
The optical properties of Xenon in the VUV, where it emits when excited by ionizing radiation, determine the performances of the calorimeter. These properties are not very well known for LXe. Measurements are difficult and experimental results are often not fully compatible. As it will be discussed in detail in the next sections, extensive measurements are available for gaseous Xe at various pressures, many in the visible region and some in the UV. The present note critically examines the experimental data. It discusses the limits in thermodynamical variables like temperature, pressure, density, etc. within which Xe can be described by a simple approximation of a more general expression valid for generic fluids. It discusses how and up to which point measurements, different from the ones in the VUV for LXe, can be used to gain insight on optical properties of interest for scintillation calorimetry.   

\section{Properties of LXe of interest for a scintillation calorimeter}
\label{sec:Xesummary}
We list in Table~\ref{tab:ypsilantis} and 
Table~\ref{tab:isotopic} quantities relative to Xenon which are of interest for its use in an experiment. Reference is made to the sources of this information.
\begin{table}
\begin{center}
\begin{minipage}{\textwidth}
\begin{tabular}{llll}
\hline
Material Properties              & Value \& Unit & Conditions  & Ref.\\
\hline
Atomic Number                    & 54            &             & \\
Atomic Weight A                  & 131.29 g/mole &             & \cite{crc}\\
Boiling point $T_b$              & 165.1~K       & 1 atm       & \cite{crc}\\
Melting point $T_m$              & 161.4~K       & 1 atm       & \cite{crc}\\
Density $\rho_{\rm liq}$         & 2.98~g/cm$^3$ & 161.35~K    & \cite{sin69}\\
Volume ratio $\rho_{\rm gas}/\rho_{\rm liq}$& 550 & 15~$^\circ$C, 1~bar& \cite{airliquide}\\
Critical point $T_c$, $P_c$      & 289.7~K, 58.4~bar &         & \cite{airliquide}\\
Triple point $T_3$, $P_3$        & 161.3~K, 0.816~bar &        & \cite{airliquide} \\
Radiation length $X_0$           & 2.77~cm       &  in liquid  & \cite{pdb}\\
                                 & 8.48~g/cm$^2$       &              & \\

Moli\`ere radius $R_M$           & 5.6~cm        &             & \cite{pdb}\\
Critical Energy                  & 10.4~MeV      &             & \cite{pdb}\\
$-({\rm d}E/{\rm d}x)_{\rm mip}$ & 1.255~MeV cm$^2$/g   &             & \cite{pdb}\\
Refractive index                 & $1.6\div1.72$ & in liquid at 178~nm & \cite{bar, chepel}\footnote{
Discrepancies are present among the measured values. Refractive index in~\cite{bar} was determined at 180~nm. } \\
Fano Factor                      & 0.041         & theoretical & \cite{doke:portugal}\\
                                 & unknown       & experimental& \\
Energy/scint.~photon $W_{\rm ph}$& $(23.7\pm2.4)$~eV  & electrons   &  \cite{doke}\\
                                 & $(19.6\pm2.0)$~eV  & $\alpha$-particles&  \cite{doke}\\
Lifetime singlet $\tau_s$        & 22~ns         &             & \cite{doke}\\
Lifetime triplet $\tau_t$        & 4.2~ns        &             & \cite{doke}\\
Recombination time $\tau_r$      & 45~ns & dominant for $e,\gamma$  & \cite{doke} \\
Peak emission wavelength $\lambda_{\rm scint}$ & 178~nm & 
& \cite{ jortner:article,jortner:book} \\
Spectral width (FWHM)            & $\sim 14$~nm  &  &  \cite{jortner:article,jortner:book} \\
Scint.~Absorption length $\lambda_{\rm abs}$ & $>100$~cm &     & \cite{proposal} \\
Rayleigh scattering length $\lambda_{\rm R}$ &$(29\pm2)$~cm    &             & \cite{ishida}\\
Thermal neutron $\sigma_{\rm tot}$  & $(23.9 \pm 1.2)$~barn& Natural composition & \cite{ncs}\\ 
\hline
\end{tabular}
\end{minipage}
\end{center}
\caption{\label{tab:ypsilantis}Main properties of liquid Xe.}
\end{table}

\begin{table}
\begin{center}
\begin{tabular}{cc}
\hline
Isotope & Abundance (\%) \\
\hline
$^{124}$Xe  & 0.096      \\
$^{126}$Xe  & 0.090      \\
$^{128}$Xe  & 1.92      \\
$^{129}$Xe  & 26.44      \\
$^{130}$Xe  & 4.08      \\
$^{131}$Xe  & 21.18      \\
$^{132}$Xe  & 26.89      \\
$^{134}$Xe  & 10.44      \\
$^{136}$Xe  & 8.87      \\
\hline
\end{tabular}
\end{center}
\caption{\label{tab:isotopic}Isotopic composition of natural Xe.}
\end{table}

LXe has interesting properties as a scintillating detector medium. It has high density and high atomic number, it competes with inorganic crystals like NaI in the number of photons emitted per MeV, it has a fast response, it allows
pulse shape analysis and particle identification. 
The light emission is in the VUV.
The scintillation mechanism, which involves excited atoms Xe$^\ast$ 
and Xe$^+$ produced by ionizing radiation, can be summarized as follows:
\begin{equation}
\begin{array}{l}
{\rm Xe}^\ast + {\rm Xe} \to {\rm Xe}_2^\ast \to 2 {\rm Xe} + h\nu
\end{array}
\end{equation}
or
\begin{equation}
\begin{array}{lllll}
{\rm Xe}^+ + {\rm Xe} \to & {\rm Xe}_2^+, & & \\
& {\rm Xe}_2^+ + e \to & {\rm Xe}~+ & {\rm Xe}^{\ast \ast}, \\
& & & {\rm Xe}^{\ast \ast} \to & {\rm Xe}^\ast + {\rm heat} \\
& & & & {\rm Xe}^\ast + {\rm Xe}\to   {\rm Xe}_2^\ast   \to 2 {\rm Xe} + h\nu,\\
\end{array}
\end{equation}
where $h \nu$ is the ultraviolet photon.
A remarkable feature of these processes is the close similarity between the emission of gaseous and condensed pure noble gases. Such a behaviour is basically due to the fact that in both cases the last relaxation step before the radiative decay is the formation of an ``excimer-like'' state. This peculiarity is a strong hint at expecting that noble liquids are transparent to their scintillation light, since the emitted VUV photons have an energy which is appreciably
 lower than the minimum one for Xe atom excitation.
Liquid rare gas scintillation detectors were the subject of recent reviews \cite{doke}. The main properties of LXe we listed in  
Table~\ref{tab:ypsilantis} are the ones
we are presently using in the Monte Carlo simulation of the MEG experiment. 

Although an appreciable number of projects is based on the future use of
 large volume LXe detectors, only a few, rather small size, LXe detectors
 were really used in experiments \cite{ber}. The actual behaviour of a
 large size LXe calorimeter, namely: the quality of its energy measurements 
and the determination of the event position and timing, directly depend on
 the transmission of the emitted photons through the detector medium. One
 can introduce the quantities $\lambda_{\rm att},\lambda_{\rm abs},
\lambda_{\rm dif}$, where $\lambda_{\rm att}$ enters the expression 
for the light attenuation in LXe, $I(x)= I_{0} e^{-x/{\lambda_{\rm att}}}$
  and the three characteristic lengths are related by:
\begin{equation}
	\frac{1}{\lambda_{\rm att}} = \frac{1}{\lambda_{\rm abs}}+\frac{1}
{\lambda_{\rm dif}}~.
\label{reciproci}
\end{equation}
$\lambda_{\rm abs}$ corresponds to a real absorption of photons, 
which are then 
 lost for good. $\lambda_{\rm dif}$ is related to photon elastic 
scattering and 
it can be assumed to coincide with $\lambda_{\rm R}$, the Rayleigh scattering 
length. Also relevant is the knowledge of the index of refraction $n$ in
 the region of Xe VUV light emission. 
The experimental knowledge of the quantities 
$\lambda_{\rm att},
\lambda_{\rm abs},\lambda_{\rm dif}$ and $n$ for pure LXe is not satisfactory.
 Measurements in the VUV are difficult and can be influenced by spurious
 systematic effects, as discussed for instance in~\cite{doke}. 
$\lambda_{\rm abs}$ has not been determined yet. A significant lower limit was
 recently established by us \cite{proposal}. Several
 measurements of $\lambda_{\rm dif}$ and $n$ are available, but they differ
 by amounts often larger than the stated errors. It should be noted that
 in the separate determination of $\lambda_{\rm abs}$ and $\lambda_{\rm 
dif}$, the
 linear dimension of the LXe cells must be at least comparable with the value for
 the $\lambda$-parameters, if one aims at obtaining reliable measurements.
 
A possible explanation of some of the discrepancies among the available
 experimental data may be connected with the degree of purity of the LXe.
 Small amounts of VUV absorbing
 substances like H$_{2}$O or O$_2$, at the level of a few parts per
 million, can dramatically change LXe optical parameters. No safe
 determination of optical parameters can leave aside monitoring the level and 
the stability of the Xe purity.

While experimental measurements of the $\lambda$'s and $n$ in LXe are scanty and suffer from problems, better information is available for gaseous Xe at various pressures both for visible and VUV light \cite{hohm,sin,bid,laporte}.

One can then examine if optical properties for LXe can be derived from this information and if the extrapolated values can be trusted. The passage from a dilute gas to gas under high pressure or to the liquid phase implies a large variation in density. When this occurs non-linear effects come normally into play due to interatomic interactions.

\section{Theoretical models for the optical properties of non-polar rare gases}
\label{sec:theory}
The link between the microscopic and the macroscopic fluid properties is provided, in our case, by the so called Clausius-Mossotti relation \cite{landau,jack}.
The Clausius-Mossotti (as a function of dielectric constant $\epsilon$) and the Lorentz-Lorenz equation (as a function of the refractive index $n$) \footnote{\quad $n^2 = \epsilon$ and $n$ real are assumed to be at $\omega$'s 
far from the Xe anomalous dispersion region.}
\begin{eqnarray}
F_{LL} \equiv \frac{n^{2}(\omega)-1}{n^{2}(\omega)+ 2} = \frac{\epsilon(\omega)-1}{\epsilon(\omega)+2} = 
\frac{4 \pi}{3} \frac{N_{A}\alpha(\omega) \rho} {M}= A(\omega) \rho_{m} ,
\label{eqn:F2}
\end{eqnarray}
where $N_{A}$ is the Avogadro's number and $M$ is the molecular weight, 
provide a general relation, valid for dilute non-polar gases, among the gas index of refraction $n(\omega)$, the molecular polarizability $\alpha(\omega)$ and the density $\rho$ (units: $\rm {g\ cm^{-3}}$) or the molar density $\rho_m=\rho/M$ (units: $\rm { mol\ cm^{-3}}$). As we shall later see, $A(\omega)$ represents ``the first virial coefficient''. 
At higher gas densities and for the liquid and solid states, the simple equations (\ref{eqn:F2}) are in general no longer valid and non-linear effects come into play due to interatomic interactions.
In this case the Clausius-Mossotti and Lorentz-Lorenz equations must be generalized into:
\begin{eqnarray}
F_{LL} \equiv \frac{n^{2}(\omega, T, \rho)-1}{n^{2}(\omega, T, \rho)+2} = \frac{\epsilon(\omega, T, \rho)-1}{\epsilon(\omega, T, \rho)+2} = 
\frac{4 \pi}{3} N_{A}\alpha(\omega, T, \rho) \rho_m ,
\label{eqn:F1}
\end{eqnarray}
and one often uses a so called ``virial expansion'' of the type:
\begin{equation}
F_{LL}=[A_{R}(\omega, T)+B_{R}(\omega, T)\rho_{m} +C_{R}(\omega, T)\rho_{m}^{2}+...] \rho_{m}
\label{eqn:F3}
\end{equation}
where the suffix $R$ indicates that $A_{R}, B_{R}, C_{R}$ are meant to be at optical and VUV wavelengths, unlike $A_{\epsilon}, B_{\epsilon}, C_{\epsilon}$ which are meant to be at the limit of  very large wavelengths. The different terms of the virial expansion can be interpreted as follows: $A_{R}$ represents the atomic polarizability, $B_{R}$ represents the interactions of two atoms, $C_{R}$ represents the interactions of three atoms, etc. 
If instead of dealing with optical properties one deals with the fluid equation of state, one often uses the ``compressibility'':
\begin{equation}
Z=\frac{P}{RT\rho_{m}},
\label{eqn:F4}
\end{equation}
where $\rho_{m}$ is the molar density. For a perfect gas $Z=1$. For a real gas $Z$ measures the deviations from the perfect gas behaviour. A virial expansion can be introduced for $Z$: 
\begin{equation}
Z=1 + B_{2V}(T)\rho_{m} + B_{3V}(T)\rho_{m}^2 + B_{4V}(T)\rho_{m}^3 + ...
\label{eqn:F5}
\end{equation}
Interatomic interactions are taken care of by the terms $B_{2V}, B_{3V}, B_{4V}$, etc. It is important to note that if the $B$'s are experimentally known for the fluid, one can use a $F_{LL}$ virial expansion in terms of the perfect gas molar density:
\begin{equation}
\rho_0=\frac{P}{RT},
\label{eqn:G6}
\end{equation}
with coefficient which can be expressed in terms of the ones of (\ref{eqn:F3}). 
The expression (\ref{eqn:F3}) becomes simpler for Xenon taking into account the fact that this gas has no permanent dipole moment.  It can be shown \cite{hohm} that this reflects into an $A_{R}(\omega,T) \rightarrow A_{R}(\omega)$, no longer depending on temperature and:
\begin{equation}
F_{LL}=[A_{R}(\omega)+B_{R}(\omega, T)\rho_{m} +C_{R}(\omega, T)\rho_{m}^{2}+...] \rho_{m}.
\label{eqn:F6}
\end{equation}

The question is: can one use for Xenon, whatever its pressure and state, the simple Clausius-Mossotti equation, so keeping only the first term of the virial expansion and a linear relation with density~?
We briefly discuss this point in section \ref{sec:light}, after critically examining in section \ref{sec:expe} the existing literature and the experimental measurements on the relevant Xenon properties. 
\section{Experimental measurements available for Xe}
\label{sec:expe}
Over the years several measurements were produced of quantities like the real and imaginary parts of the dielectric constant ($\epsilon_1,\epsilon_2$), the real and the imaginary part of the refractive index ($n,k$), the molecular polarizability $\alpha(\omega,T,\rho)$, etc. over a large range of temperatures T and pressures P, from dilute gas to liquid and solid phases.
We present and discuss the available data and papers.
\subsection{Experimental determination of the dielectric virial coefficients of atomic gases as a function of temperature. J.\ Hout and T.K.\ Bose \cite{huot}} 
State: gas.
Temperature: $T=323.15$~K.
Density: $0.13 < \rho < 0.79~{\rm g\ cm^{-3}}$ equivalent to $10^{-3} < \rho_m < 6.0 \ 
10^{-3}~{\rm mol\ cm^{-3}}$.
Method: Absolute measurement of the dielectric constant in capacitances (at low frequency).
Result: Determination of the first three virial coefficients $A_{\epsilon}$, $B_{\epsilon}$, $C_{\epsilon}$. At 323.15 K $A_{\epsilon}=10.122 \pm 0.002 ~{\rm cm^3\ mol^{-1}}$, $B_{\epsilon}=32 \pm 2 ~{\rm cm^6\ {mol^{-2}}}$, $C_{\epsilon}=-3482 \pm 310 ~{\rm cm^9\ {mol^{-3}}}$.
This measurement is made at very low frequency, so very far from the visible and the VUV region, and at densities which are about one hundred times smaller than the one of LXe. Moreover, models were proposed for the dependence of $B_{\epsilon}$ and $C_{\epsilon}$ on the temperature $T$, which seem not to be confirmed by experimental data. It is remarkable that, if one, somewhat arbitrarily, computes the Clausius-Mossotti relation in the visible and at densities up to a density $\rho_m= 0.016~{\rm mol\ cm^3}$  using $A_{\epsilon}, B_{\epsilon}, C_{\epsilon}$ of this paper, one finds that $A$ is not very different from direct determinations closer to the specified region \cite{acht} and that $B_{\epsilon}$ and $C_{\epsilon}$ introduce terms which produce rather small deviations from the main $A$ term. 

\subsection{Experimental determination of the refractivity virial coefficients of atomic gases. H.J.\ Achtermann et al. \cite{acht}} 
State: gas.
Temperature: $T=298-348$~K.
Pressure: $p < 4\times10^2$~Atm.
Density:  $\rho_m < 0.016~{\rm mol\ cm^{-3}}$.
Method: Measurement of the refractive index by a differential interferometric technique at $\lambda=633$~nm.
Result: Determination of the first three virial coefficients $A_{R}, B_{R}, C_{R}$. $A_{R}=10.344 \pm 0.002~{\rm cm^{3}\ mol^{-1}}$, $B_{R}=28.5 \pm 0.5~{\rm cm^{6}\ mol^{-2}}$, $C_{R}=-1802.0 \pm 50.0~{\rm cm^{9}\ mol^{-3}}$ at $T=348$~K.
 The measurements were made in the visible region at densities close to the one of LXe. One observes that $A_{R}$ is not very different from $A_{\epsilon}$ of the static measurement. The terms related to $B_{R}$ and $C_{R}$ are small in respect of the main $A_{R}$ one at LXe densities.  

\subsection{Refractive indices of the condensed rare gases, argon, krypton and xenon. A.C.\ Sinnock \cite{sin}} 
State: liquid and solid Xe.
Temperature: $T=30-178$~K.
Pressure: $p \approx 1$~Atm.
Density:  $\rho_m = 0.0215-0.0280~{\rm mol\ cm^{-3}}$.
Method: Measurement of the refractive index by a total reflection refractometer at $\lambda=546.1$~nm.
It studies Xe in the liquid and solid forms and in the visible region.
Result: Determination of the refractive index and determination of $A$ from the Lorentz-Lorenz function $F_{LL}$. In LXe  $n \approx 1.4$ and $A \approx 10.5 \pm 0.025 ~{\rm cm^3\ mol^{-1}}$ at $\lambda=546.1$~nm. The value of A is remarkably similar to the ones in gas over a large density range.
In a previous article \cite{sin69}, similar measurements were performed also at different wavelengths ($361.2 < \lambda < 643.9$~nm). The Lorentz-Lorenz function, hence $A$, only varies $\approx 3 \%$ over the entire range studied; this implies that the variation of $n$ is $\approx 1.5 \%$ in the specified region.
\subsection{Frequency-dependence of second refractivity virial coefficient of small molecules between 325 nm and 633 nm. U.\ Hohm \cite{hohm1}} 
State: gas.
Temperature: Room temperature. 
Pressure: $p < 3$~Atm.
Density:  $\rho_m < 1.25 \ 10^{-4}~{\rm 
mol\ cm^3}$.
Method: Measurement of the refractive index $n$ as a function of the gas density by a Michelson interferometer between $325 < \lambda < 633$~nm.
Results: The first two virial coefficient $A_{R}, B_{R}$ are derived from the measurements. $A_{R}(633)= 10.43, A_{R}(594)=10.46, A_{R}(543.5)=10.52, A_{R}(325.1)=11.26~{\rm cm^3\ mol^{-1}}$.
$ 23.2 <B_{R} < 25.5~{\rm cm^{6}\ mol^{-2}}$. Apart from the normal dispersion as a function of the wavelength, the values of $A_{R}$ and $B_{R}$ are remarkably stable whatever the Xe physical state.

\subsection{Measurement of refractive indices of Neon, Argon, Krypton and Xenon in the 253.7-140.4~nm wavelength range. Dispersion relations and estimated oscillator strengths of the resonance lines. A.\ Bideau-Mehu et al. \cite{bid}} 
State: gas.
Temperature: $T=290$~K.
Pressure: $p \approx 1$~Atm.
Density:  $\rho_m = 4.49  \ {-5}~{\rm mol\ cm^{-3}}$
Method: Measurement of the refractive index by a Fabry-Perot interferometer in the region $140.4 < \lambda < 253.7$~nm.
One of the few precise measurement of the refractive index in the VUV and close to the first Xe absorption line.
We saw that in the visible region $A(\omega)$ is rather stable around $10.5~{\rm cm^3\ mol^{-1}}$ even for large density variations from dilute gas to the liquid and solid states.
One can expect that in the VUV the same is true, although normal dispersion from the visible to the VUV region corresponds to a higher value for $A(\omega)$. From these measurements of the refractive index one can derive a $A(\omega) \approx 17.0~{\rm cm^3\  mol^{-1}}$ at 
$175$~nm. The LXe refractive index, in the approximation of a Lorentz-Lorenz function linear in the density $\rho$, is predicted to be $n \approx 1.69$ at $175$~nm. 
\begin{figure}[t]
	\begin{center}
  \begin{tabular}{cc}
   \includegraphics[width=0.45\columnwidth]{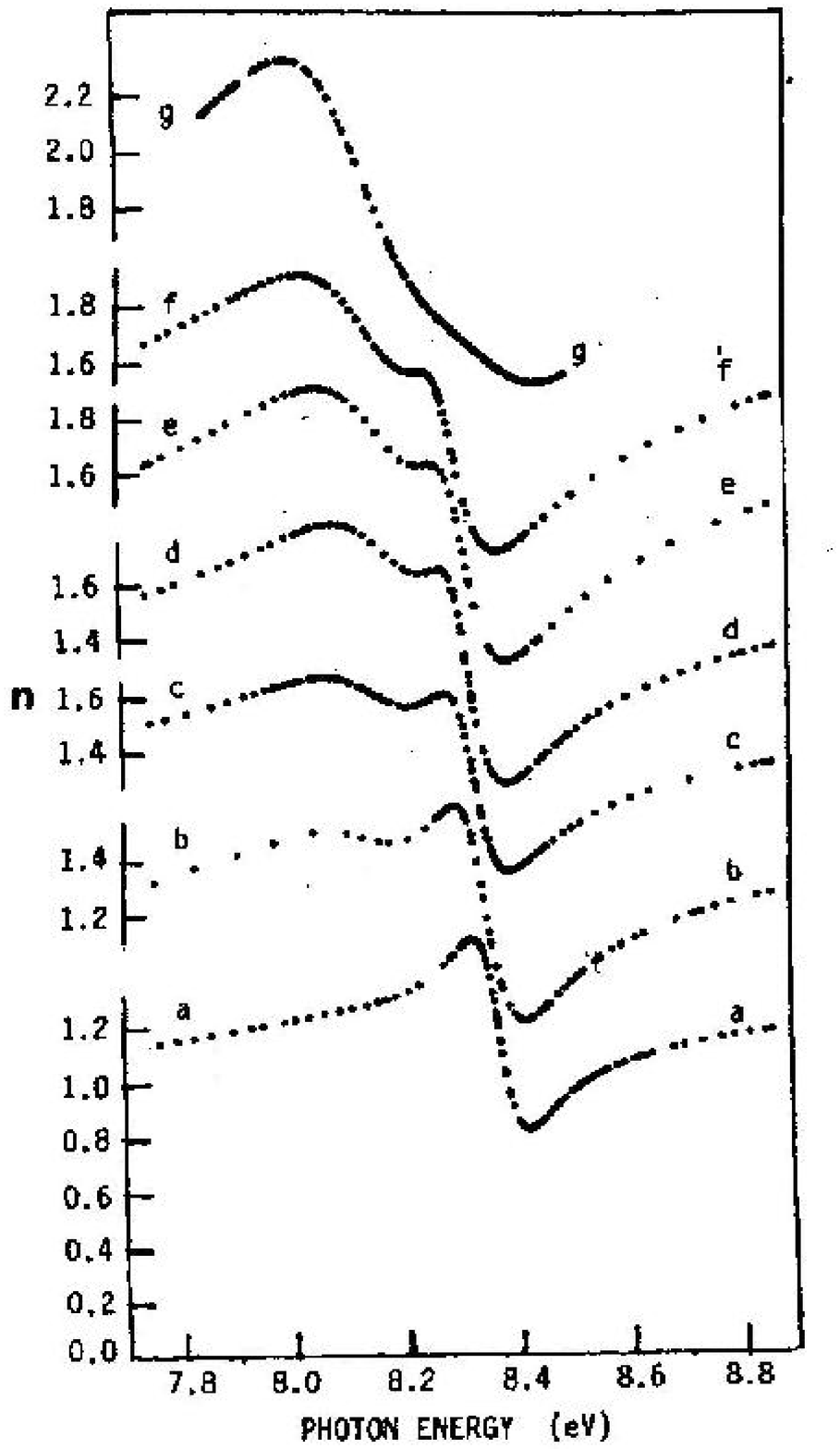} &
   \includegraphics[width=0.48\columnwidth]{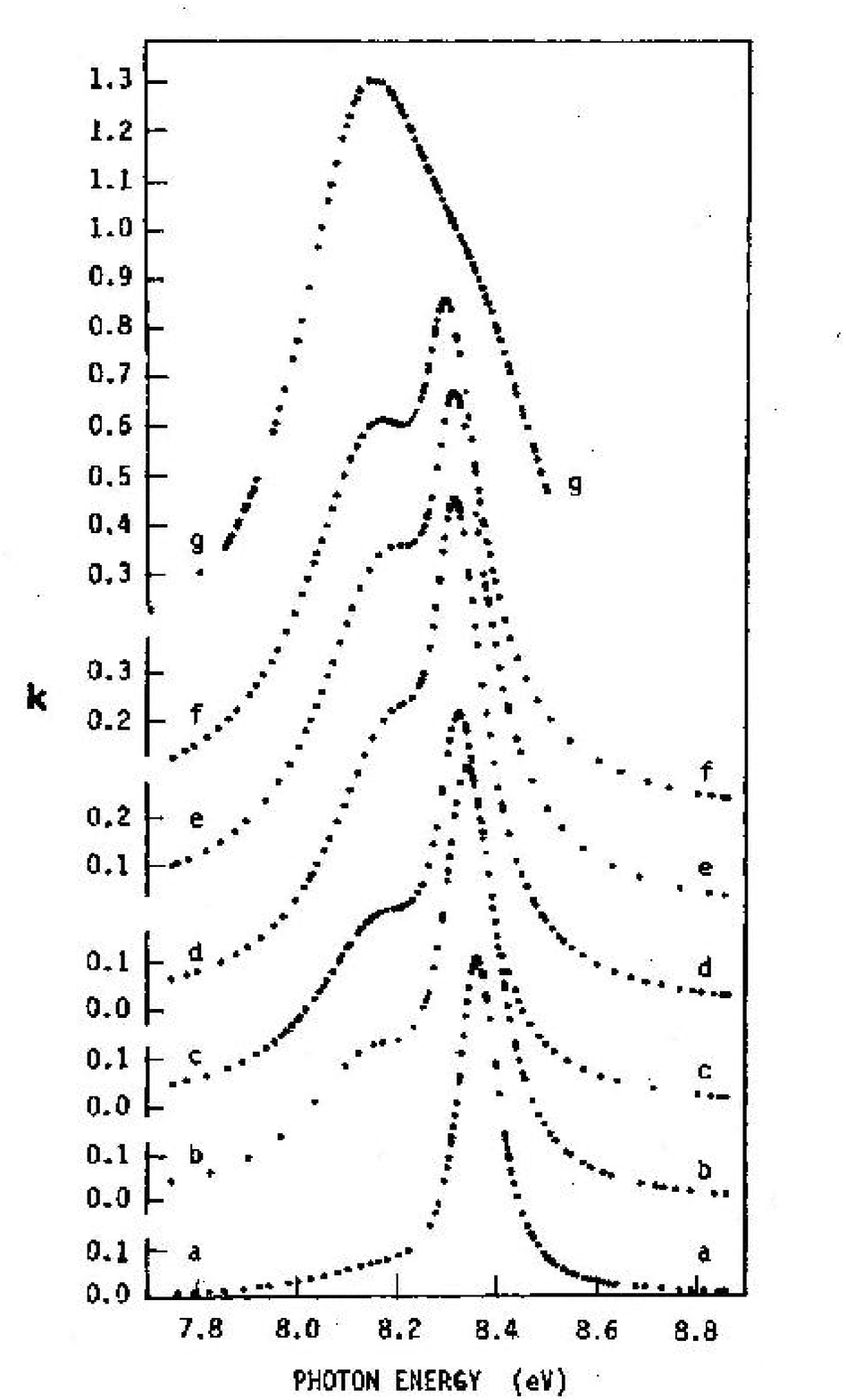} \\
  \end{tabular}
  \end{center}
\caption{The real and the imaginary parts of the Xe refractive index vs photon energy in the region of the first Xe absorption line. Increasing densities from $a$ to $g$ (the LXe emission is at 7.1 eV) ~\cite{laporte}.}
\label{fig:la1}
\end{figure}
\subsection{Measurement of refractive indices of liquid Xenon for intrinsic scintillation light. L.M.\ Barkov \cite{bar}} 
State: liquid Xenon.
Temperature: $T=161.36$~K.
Pressure: $p \approx 0.81$~Atm.
Density:  $\rho_m \approx 2.26 \  10^{-2}~{\rm mol\ cm^{-3}}$.
Method: Measurement of the refractive index by modification of the focalization of a lens when immersed in LXe, at $\lambda = 180$~nm.
Result: $n=1.5655 \pm 0.0024 \pm 0.0078$. The quoted errors are probably underestimated.
\subsection{Evolution of excitonic bands in fluid xenon. P.\ Laporte and I.T.\ Steinberger \cite{laporte}} 
State: gas under pressure and liquid Xenon.
Temperature: gas $275 < T < 292.7 $~K,  liquid $T=163$~K.
Pressure: gas $61.0 < p < 82.3$~Atm,  liquid $p=0.79$~Atm.
Density: gas $7.6 \ 10^{-3} < \rho_m < 1.45 \  10^{-2}~{\rm mol\ cm^{-3}}$, liquid $\rho_m = 2.26 \ 10^{-2}~{\rm mol\ cm^{-3}}$.
Method: Study of the real and imaginary parts of the refractive index by the reflection method, in the VUV wavelength region $115 < \lambda < 165$~nm.
Result: An important and precise study of the characteristics of the first Xe absorption lines and of their evolution as a function of the density.
The appearance of new lines and the broadening and displacement of normal lines is interpreted as due to Wannier excitons (see at section \ref{sec:light}). This can be clearly seen in fig.~\ref{fig:la1}  which shows the real and the imaginary parts of the  Xe refractive index as a function of the photon energy and (from bottom to top) for increasing densities (the curves labelled from $a$ to $f$; the curve $g$ is for LXe).
The real part of $n$ in the region of LXe emission at $\lambda=175$~nm cannot be easily extrapolated from these measurements, since one should know the complex $n$ both at lower and higher wavelengths. The data are of a sufficiently good quality to allow a test of the linearity on $\rho$ of the Lorentz-Lorenz function on a large range of densities (but at smaller wavelengths than the one of LXe emission).  

\begin{figure}[t]
	\begin{center}
		\includegraphics[scale=0.5]{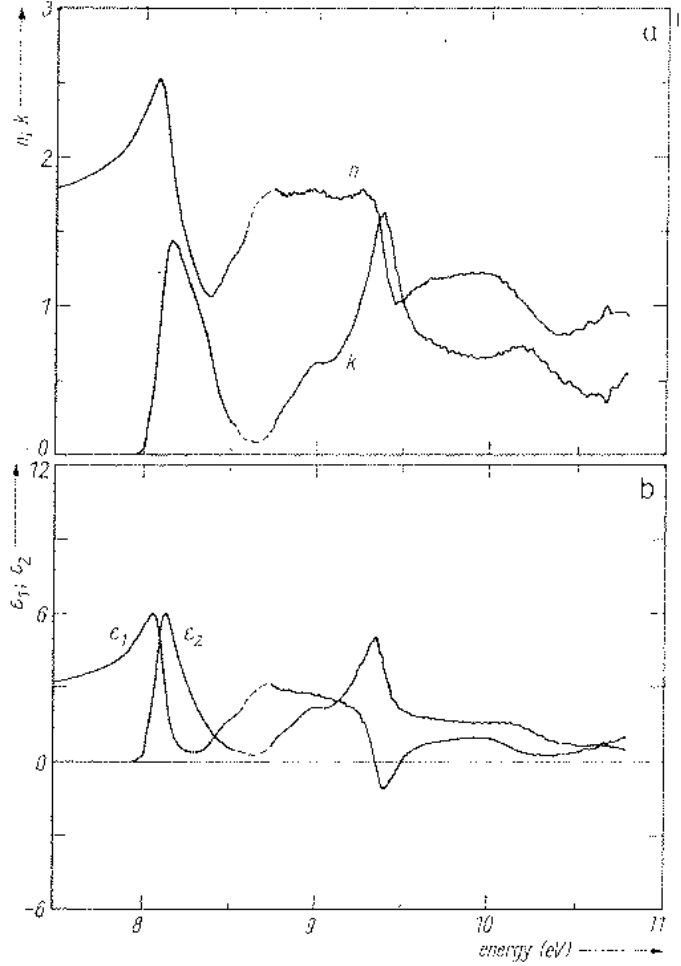}
	\end{center}
\caption{The real and the imaginary parts of the Xe refractive index and of dielectric constant vs photon energy in LXe ~\cite{subtil}.}
\label{fig:sub1}
\end{figure}

\subsection{VUV Optical Constants of Liquid and Solid Xenon at the Triple Point. J.L.\ Subtil et al. \cite{subtil}} 
State: liquid and solid Xenon.
Temperature: $161.2 < T < 163.2 $~K.
Pressure: $p \approx 1$~Atm.
Density: $\rho_m = 2.27 \  10^{-2}~{\rm mol\ cm^{-3}}$.
Method: Study of the real and imaginary parts of the refractive index by the reflection method, in the VUV wavelength region $112 < \lambda < 203$~nm.
Result: An important and precise study of the characteristics of the first Xe absorption lines and of their evolution in the passage from the liquid to the solid state. A larger range of wavelengths was studied; the LXe emission wavelength $\lambda=175$~nm ($7.1$~eV) is therefore included and the corresponding LXe refractive index is $n=1.71$. It is interesting to note that  $n=1.69$ is the value extrapolated from the VUV gas measurements of Bideau-Mehu \cite{bid}, assuming a linearity of the Lorentz-Lorenz function with the density. The passage from liquid to solid Xe in the absorption region shows  effects due to the Wannier excitons (see at section \ref{sec:light}). The real and the imaginary parts of the refractive index in LXe are shown in fig.~\ref{fig:sub1} over a wide range of photon energies. 

\subsection{Liquid Xenon Scintillation: Light Propagation and Detection. V.\ Chepel \cite{chepel}} 
State: liquid Xenon.
Temperature: $T=161.36$~K.
Pressure: $p \approx 1$~Atm.
Density: $\rho_m = 2.26 \ 10^{-2}~{\rm mol\ cm^{-3}}$.
Method: Measurement of the index of refraction $n$ and of the attenuation length $\lambda_{\rm att}$ at the emission wavelength $\lambda=178$~nm by the study of light propagation in LXe over a variable path of $\approx 10$~cm. Result: $n=1.69 \pm 0.02$ and $\lambda_{\rm att}=36.4 \pm 1.8$~cm.

\section{VUV light propagation in Liquid Xenon}
\label{sec:light}

\begin{figure}[p]
	\begin{center}
		\includegraphics[scale=0.48]{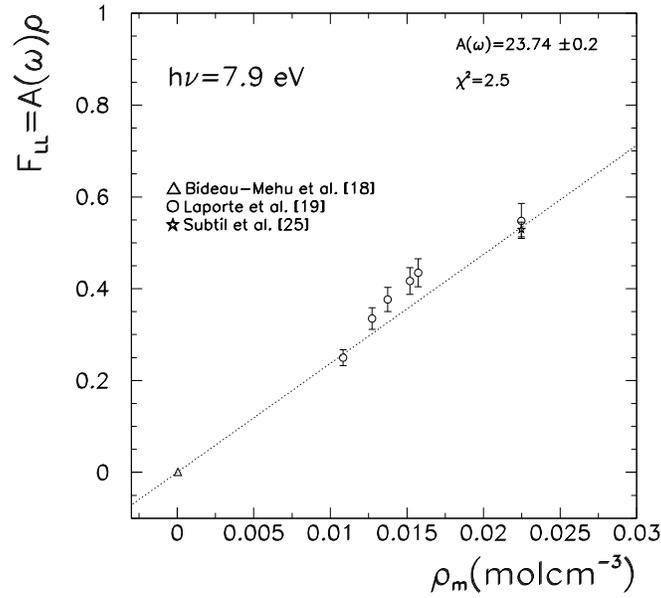}
	\end{center}
\vspace{-1cm}
\caption{Check of the linearity of $F_{LL}$ as a function of the density $\rho_{m}$, from the gas to the liquid phase at h$\nu$=7.9 eV.}
\label{fig:pap1}
\end{figure}

\begin{figure}[p]
	\begin{center}
		\includegraphics[scale=0.48]{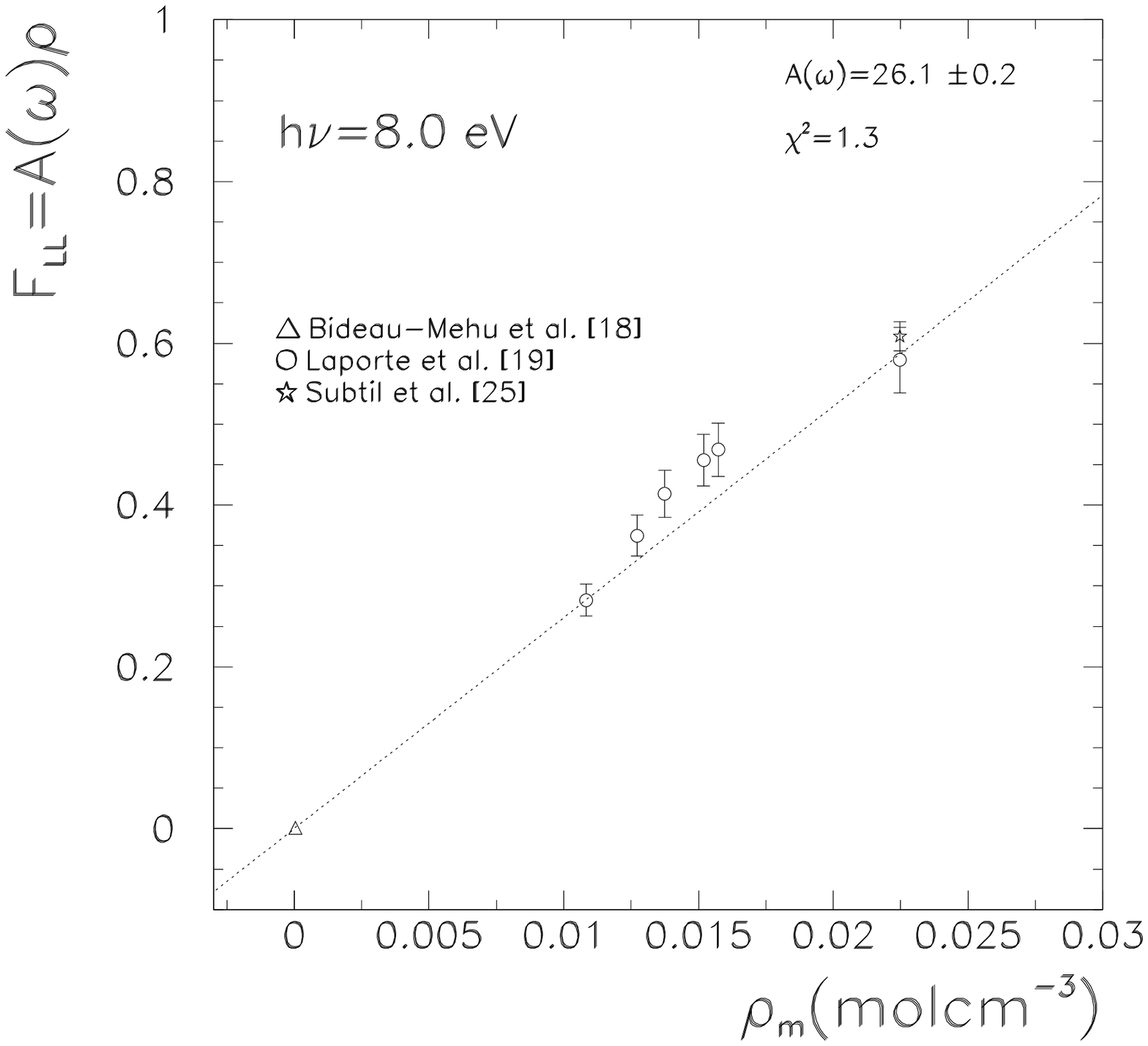}
	\end{center}
\vspace{-1cm}
\caption{Check of the linearity of $F_{LL}$ as a function of the density $\rho_{m}$, from the gas to the liquid phase h$\nu$=8.0 eV.}
\label{fig:pap2}
\end{figure}

\begin{figure}[p]
	\begin{center}
		\includegraphics[scale=0.48]{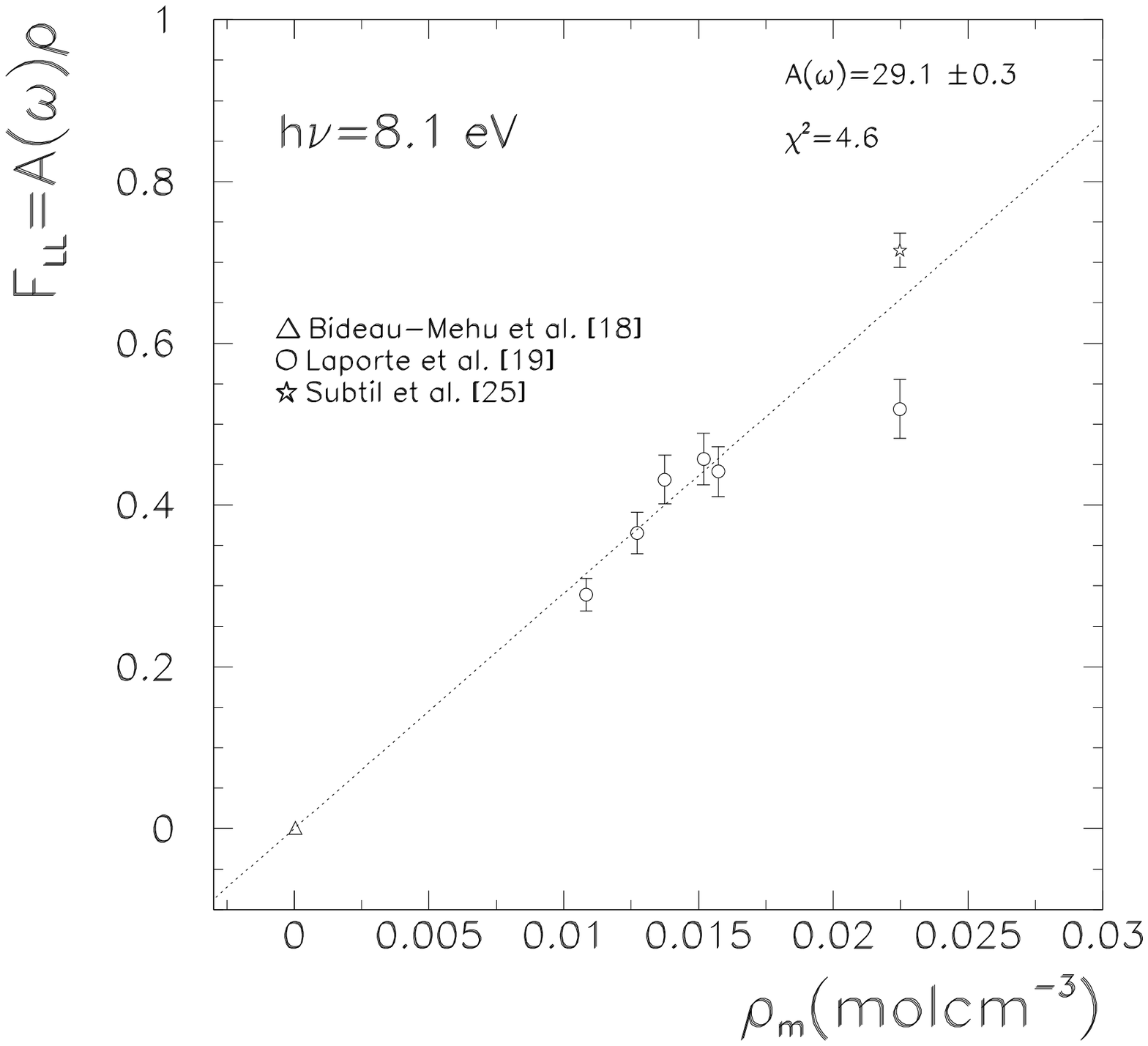}
	\end{center}
\vspace{-1cm}
\caption{Check of the linearity of $F_{LL}$ as a function of the density $\rho_{m}$, from the gas to the liquid phase h$\nu$=8.1 eV.}
\label{fig:pap5}
\end{figure}

\begin{figure}[p]
	\begin{center}
		\includegraphics[scale=0.48]{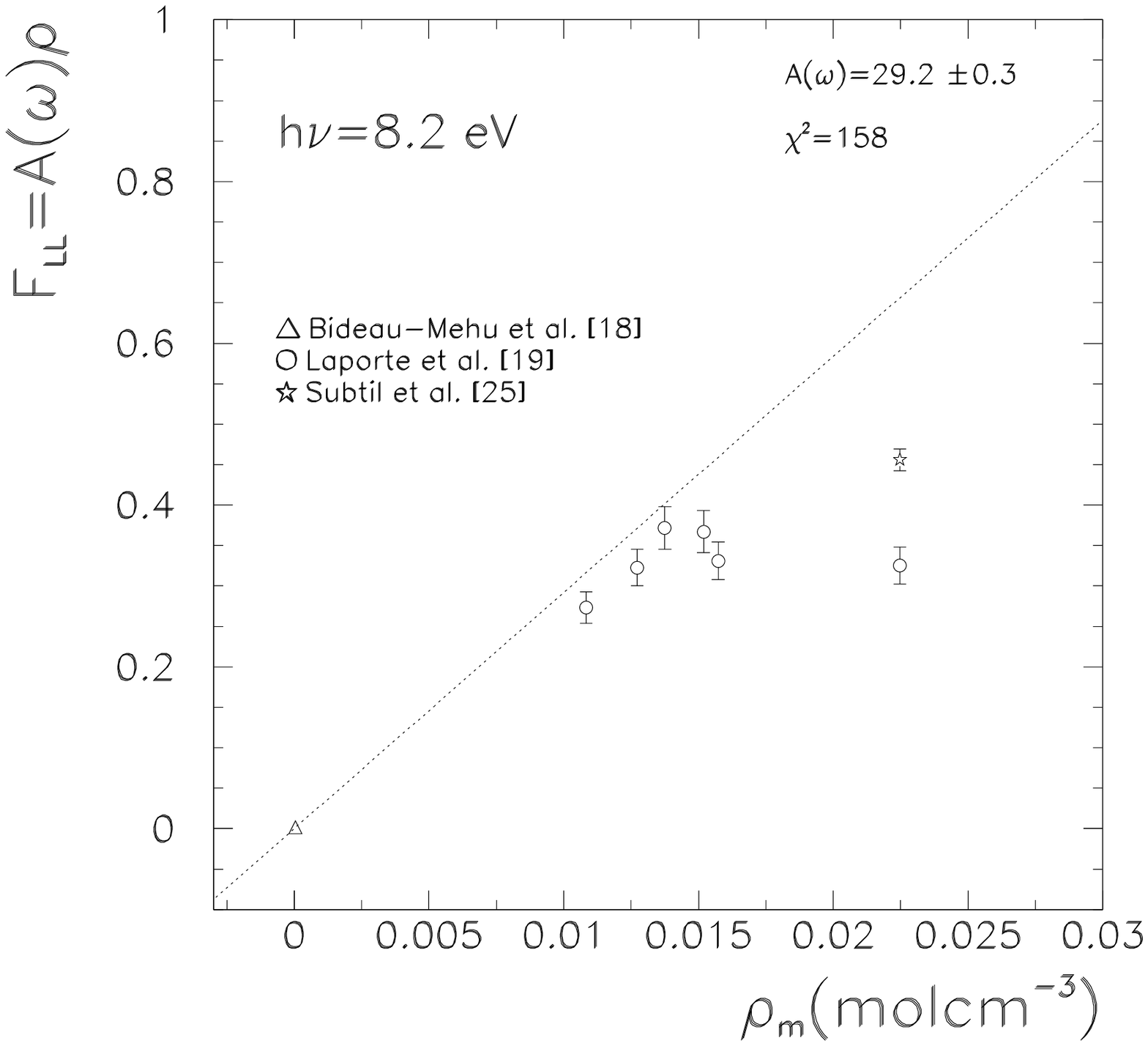}
	\end{center}
\vspace{-1cm}
\caption{Check of the linearity of $F_{LL}$ as a function of the density $\rho_{m}$, from the gas to the liquid phase h$\nu$=8.2 eV. }
\label{fig:pap6}
\end{figure}

\begin{figure}[p]
	\begin{center}
		\includegraphics[scale=0.5]{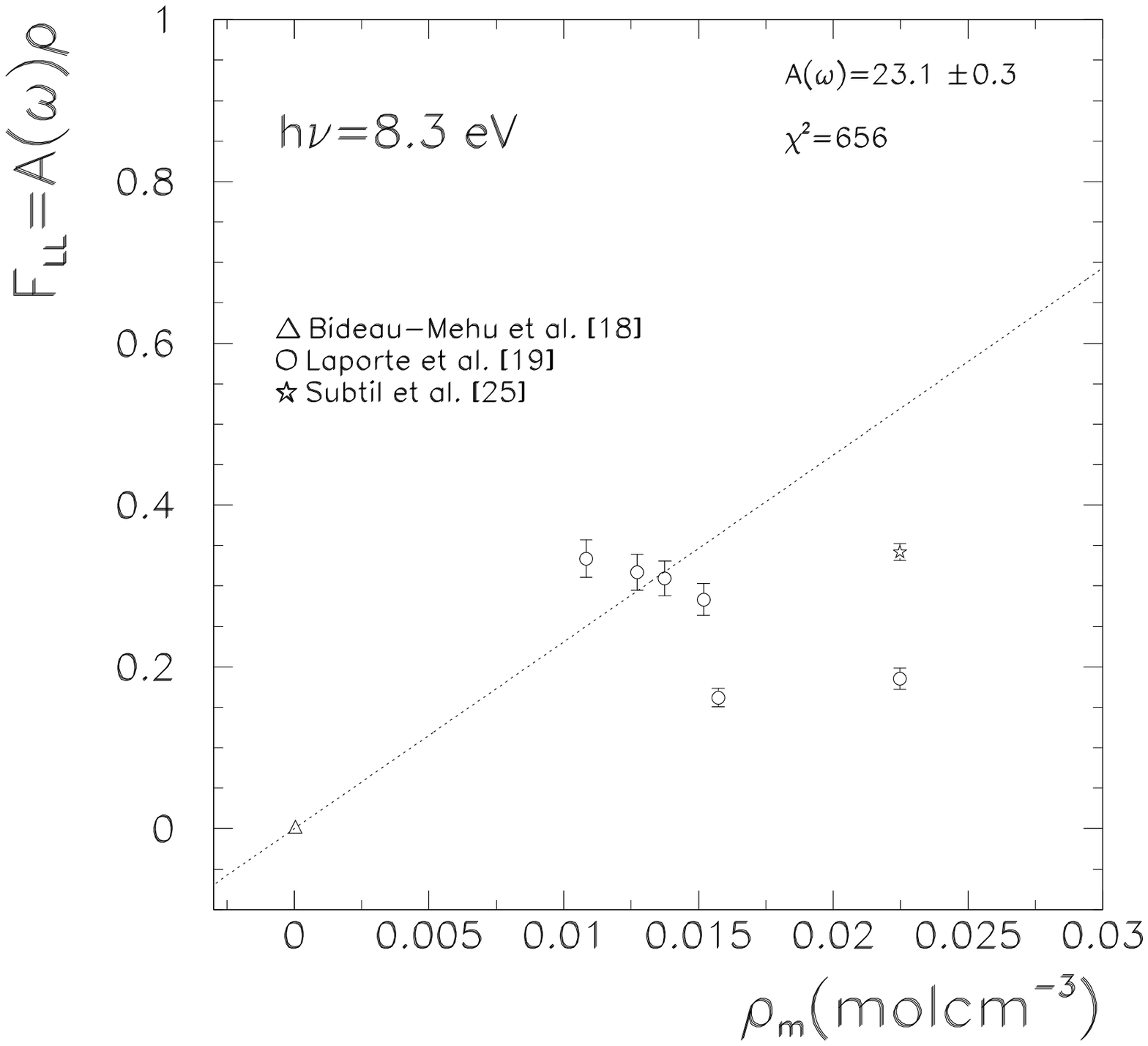}
	\end{center}
\vspace{-0.8cm}
\caption{Check of the linearity of $F_{LL}$ as a function of the density $\rho_{m}$, from the gas to the liquid phase h$\nu$=8.3 eV.}
\label{fig:pap7}
\end{figure}

\begin{figure}[p]
	\begin{center}
		\includegraphics[scale=0.75]{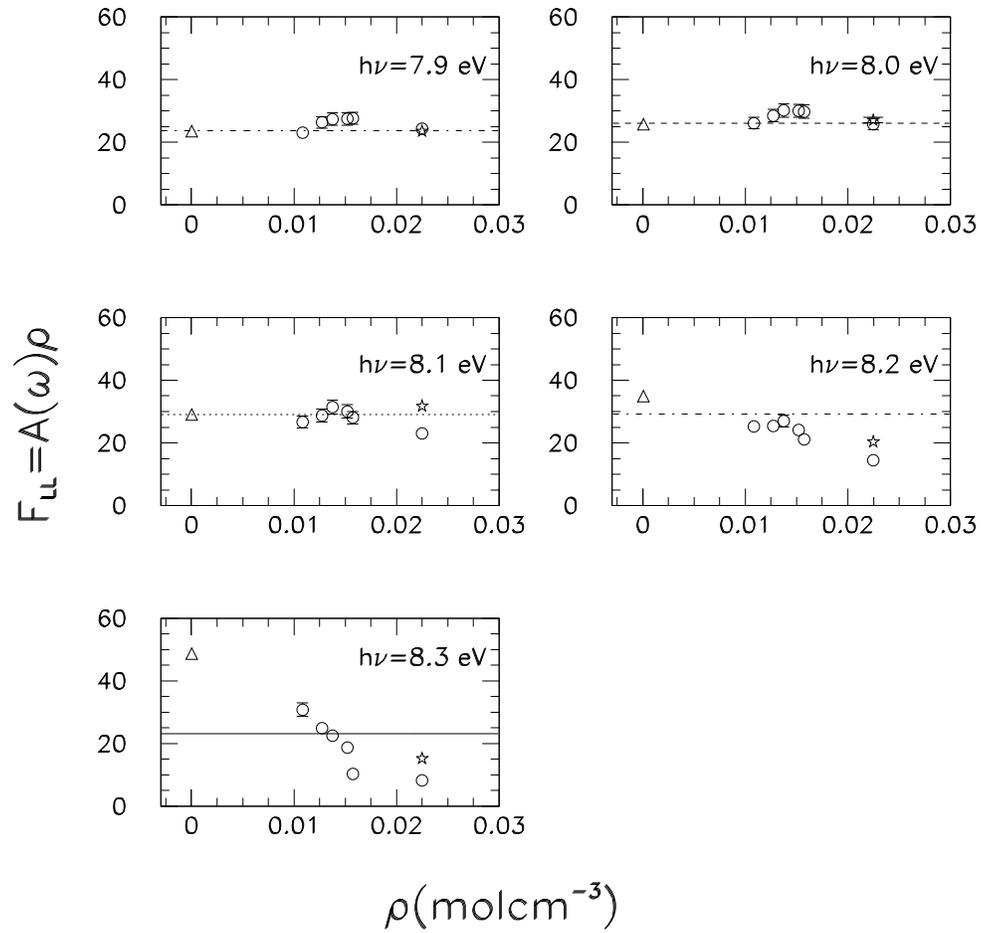}
	\end{center}
\vspace{-0.5cm}
\caption{$A(\omega)$ as a function of the density $\rho_{m}$, at different photon energies. $A(\omega)$ approximately corresponds to a constant value up to $h \nu \approx 8.1$~eV.}
\label{fig:pap10}
\end{figure}

The observation that the simple Clausius-Mossotti equation (depending only on the first virial coefficient $A_{R}$, whatever the fluid density and state) might be valid for noble gases has already been made \cite{lan}.
At a closer look for Xe, it comes out that when one is approaching the region of the Xe photon absorption lines, the Clausius-Mossotti equation is only marginally valid, but it can be considered an acceptable approximation still at the LXe emission line.  
Xenon excited by ionizing radiation emits at $\lambda=178$~nm, that is photons with $\hbar \omega=7.0$~eV . Photon absorption lines are at higher energies starting from about 8.4 eV. The index of refraction $n$ and its absorbitive part $k$ were extensively studied in the energy region from 7.8 to 8.8 eV over a wide pressure range, from dilute gas to the liquid phase \cite{laporte}. It was observed that, when the density reaches high values, the absorption lines are distorted and broadened and a system of satellite lines appears, always at energies lower than the main lines. This phenomenon is explained in terms of exciton formation \footnote{by Wannier-excitons one means Rydberg levels, at energies between the valence band and the conduction band, due to electron-hole bound states extending over $\approx10~$nm regions of the fluid.} \cite{laporte,rice,steinberger} and can be immediately related to higher density terms in the virial expansion. As a consequence the Clausius-Mossotti  relation is no longer linear in $\rho_m$ at photon energies close to the absorption lines.
Nevertheless we checked that, when one moves to lower energies, the linearity in $\rho_m$ appears to be recovered and the values of $\alpha(\omega, T,\rho_m)$ in (\ref{eqn:F1}) come out to be roughly the same, whatever the pressure. This means that in the virial expansion:
\begin{equation}
F_{LL}=[A_{R}(\omega)+B_{R}(\omega, T)\rho_{m} +C_{R}(\omega, T)\rho_{m}^{2}+...] \rho_{m}
\label{eqn:F7}
\end{equation}
the higher order terms give small contributions and that the $A_{R}$ term dominates. 
Since, as already noted, some discrepancies exist among the optical data obtained by different authors, in LXe at 7.0~eV,  no fully reliable value of optical parameters at 7.0 eV is available or can be extrapolated from higher energy measurements. The only conclusion we can reliably reach is that, in the energy region around 7.0 eV, the simple Clausius-Mossotti equation is valid with good approximation and that this statement rests on experimental data. This is shown in fig.~\ref{fig:pap1}, \ref{fig:pap2}, \ref{fig:pap5}, \ref{fig:pap6} and \ref{fig:pap7} where all the VUV results~\cite{bid,laporte,subtil} were used to compute $F_{LL}=A(\omega)\rho_m$, in the hypothesis of the validity of the simple Clausius-Mossotti equation. 
One can observe that for energies up to $8.1$~eV the $F_{LL}$ data are well fitted by a linear dependence on the density, over quite a wide density range from dilute gas to liquid. $A(\omega)$, the slope of the fitted lines, corresponds to a rather constant value, up to 8.1~eV, as it appears in fig.~\ref{fig:pap10}. At higher photon energies, hence closer to the Xe photon absorption lines, the fits become increasingly worse, the linear model is rejected (and the error on $A(\omega)$ is meaningless). 

\newpage

\begin{figure}[p]
	\begin{center}
		\includegraphics[scale=0.7]{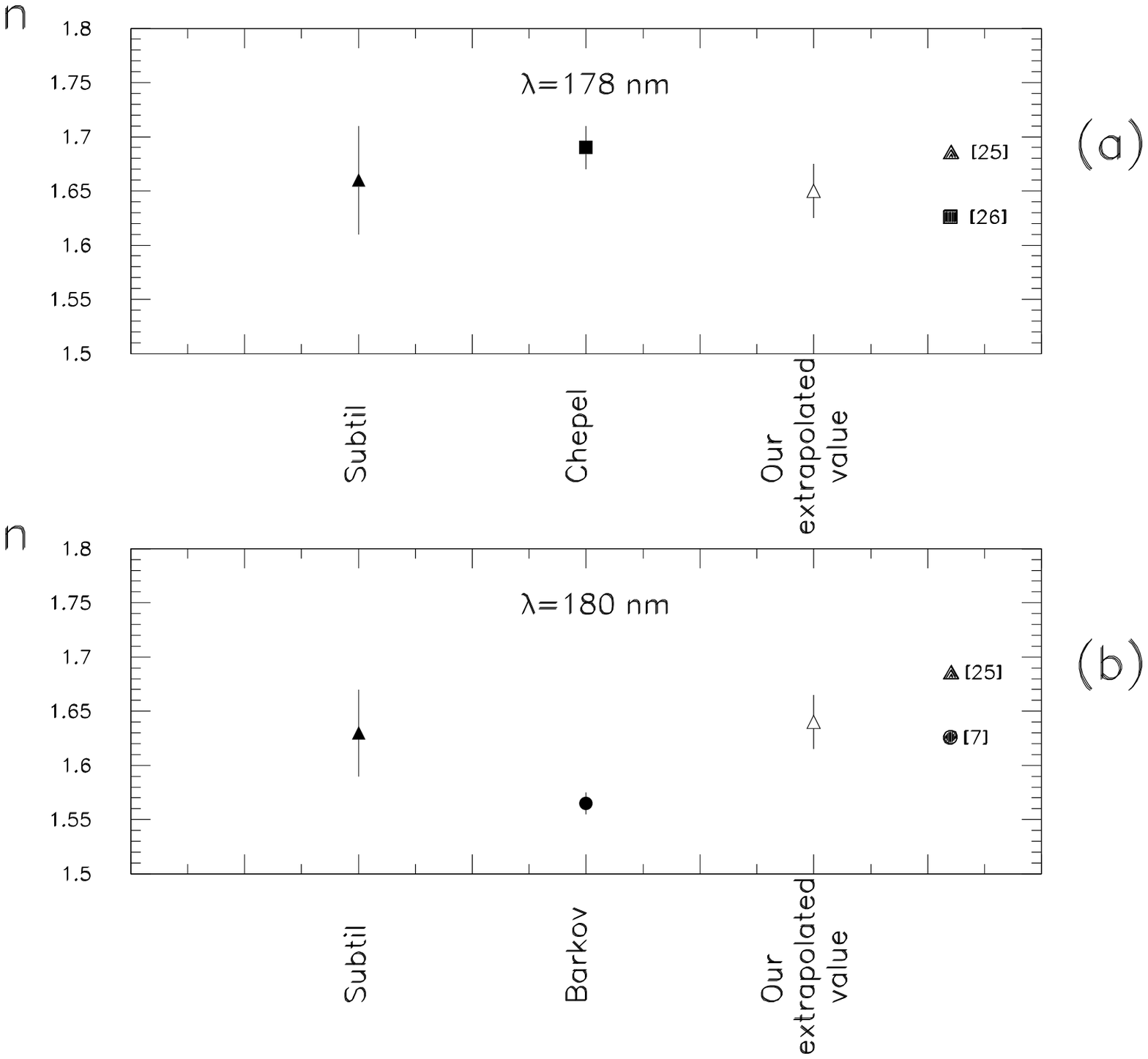}
	\end{center}
\vspace{-0.5cm}
\caption{Measurements of the refractive index in LXe (black symbols) are compared with a value extrapolated from low densities (open triangle), at two different wavelenghts.}
\label{fig:pap3}
\end{figure}

One can therefore conclude that, when $\omega$ is reasonably far from the absorption region, most of the information obtained for gaseous Xenon in various physical conditions can be used for predicting the optical properties of LXe at the emission wavelength \cite{lan}. The experimental determinations of the refractive index in LXe, at the Xe emission wavelength of $178$~nm \cite{subtil,chepel} and at $180$~nm \cite{bar}, are presented in fig.\ref{fig:pap3} (a) and (b). The errors on $n$ are the ones quoted in the papers with the exception of \cite{subtil}, which has very small errors from point to point, but a systematic error on $n$ which is not clearly stated; we assumed a systematic error of $\approx3\%$. The measured refractive index is compared with an extrapolated value we derived, based on the simple Clausius-Mossotti equation utilizing VUV data obtained in gas at low density \cite{bid}. One can observe a rather good agreement between the measured $n$ and the extrapolated value. The measurement corresponding to \cite{bar} gives a $n$ value at 180 nm which is somewhat low, but the experimental error, quoted in the paper, appears to us over-optimistic.

\begin{figure}[p]
	\begin{center}
		\includegraphics[scale=0.7]{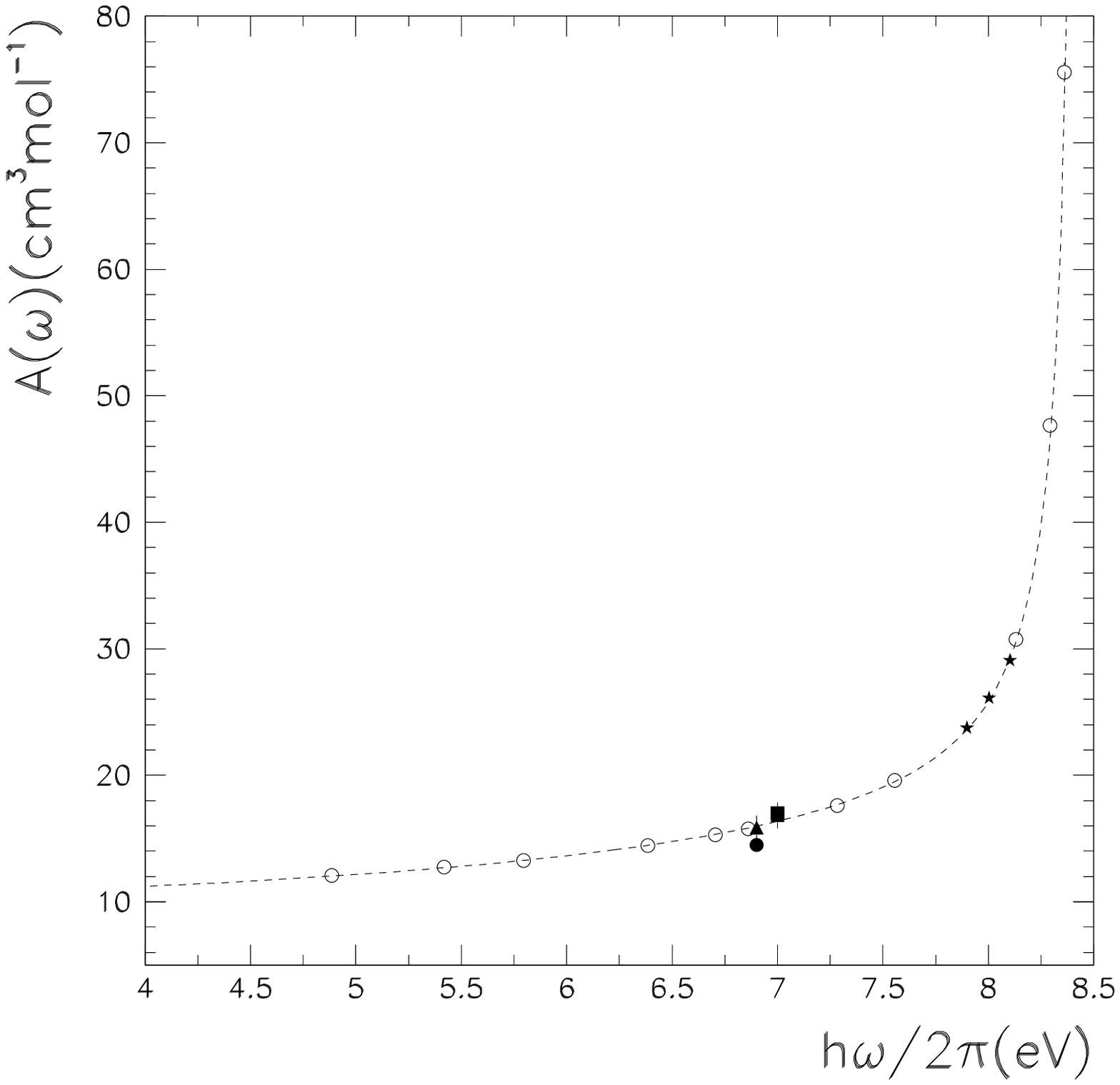}
	\end{center}
\vspace{-0.5cm}
\caption{The dispersive behaviour of $A(\omega)$ as a function of the photon energy $\hbar \omega$. Experimental measurements at higher densities are also shown (see text). }
\label{fig:pap4}
\end{figure}

Since we assumed (and checked) the validity of the simple Clausius-Mossotti equation, the relation between $n(\omega)$ and $A(\omega)$ is fixed. We present in fig.\ref{fig:pap4} the dispersive behaviour of $A(\omega)$ as a function of the photon energy $\hbar \omega$. The open circles are the VUV measurements \cite{bid} obtained at low density ($p \approx 1~{\rm Atm}$).  The dashed line is the fit to those points. The function used is: 
\begin{equation}
	A(E)\approx\sum_{i=1}^{3}\frac{P_{i}}{E^{2}-E_{i}^{2}}
\label{eqn:fit}
\end{equation}

with $E=\hbar\omega$, $P_{i}$ the parameters resulting from the fit and $E_{i}$ the absorption line positions ($P_{1}=71.23$ ${\rm eV^{2}\ cm^{3}\ mol^{-1}}$, $P_{2}=77.75$ ${\rm eV^{2}\ cm^{3}\ mol^{-1}}$, $P_{3}=1384.89$ ${\rm eV^{2}\ cm^{3}\ mol^{-1}}$, $E_{1}=8.4$ eV $E_{2}=8.81$ eV and $E_{3}=13.2$ eV).  
The star symbols are the values of $A(\omega)$ corresponding to the fits of fig.~\ref{fig:pap1}, \ref{fig:pap2}, obtained over the large density range from dilute gas to liquid Xe. The experimental measurements of the refractive index $n$ in LXe at $178$ and $180~{\rm nm}$ \cite{bar,subtil,chepel} are also shown by their corresponding $A(\omega)$, as black symbols.   

\section{Relation between the Rayleigh scattering length and the refractive index}
\label{sec:scattering}
We discuss now the relation between the index of refraction $n$ and the Rayleigh scattering length $\lambda_R$. For dilute Xe a simple relation is valid: \cite{jack}:
\begin{equation}
\frac{1}{\lambda_{R}} \simeq \frac{2 k^{4}}{3 \pi N}\left|(n-1)\right|^{2}
\label{eqn:hR}
\end{equation}
where $k$ is the wave number of the radiation and $N$ is the number of molecules per unit volume.
For  a dense fluid like LXe $\lambda_R$ depends on density and temperature fluctuations of the medium, according to the Einstein's expression ~\cite{landau}:
\begin{equation}
    \frac{1}{\lambda_{R}} = \frac{\omega^{4}}{6 \pi c^{4}}\left[K T \rho^{2} \kappa_{T}\left(\frac{\partial \epsilon}{\partial \rho}\right)^{2}_{T}+\frac{K T^{2}}{\rho c_{v}}\left(\frac{\partial \epsilon}{\partial T}\right)^{2}_{\rho}\right]
\label{eqn:h1}
\end{equation}
where $c$ is the speed of light, $\kappa_{T}$ is the isothermal compressibility, $c_{v}$ is the specific heat at constant volume and $K$ is the Boltzmann's constant. This expression reduces to (\ref{eqn:hR}) in the case of dilute gases.\\
Since Xenon is a non-polar fluid, the second part of (\ref{eqn:h1}) comes out to be neglegible ~\cite{lan,hohm,sin69}. The derivative appearing in the first part of (\ref{eqn:h1}) can be computed from the Clausius-Mossotti equation (\ref{eqn:F2}).
The Einstein's equation reduces to:
\begin{eqnarray}
	\frac{1}{\lambda_{R}} &=& \frac{\omega^{4}}{6 \pi c^{4}}\left\{K T  \kappa_{T}\frac{[n^2(\omega)-1]^{2}[n^2(\omega)+2]^{2}}{9}\right\}. \nonumber \\
\label{eqn:h3}
\end{eqnarray}	
This equation establishes a useful relation between the index of refraction in  pure LXe and the Rayleigh scattering length, which can undergo an experimental test \footnote{To the best of our knowledge this relation was never previously applied to LXe.}. Due to the form of the relation and to error propagation, for a $n \approx 1.65$ in LXe at $\lambda \approx 178$~nm $\frac{\sigma_{\lambda_{R}}}{\lambda_{R}} \approx 8.6 \frac{\sigma_{n}}{n}$.

\begin{figure}[pt]
	\begin{center}
		\includegraphics[scale=0.8]{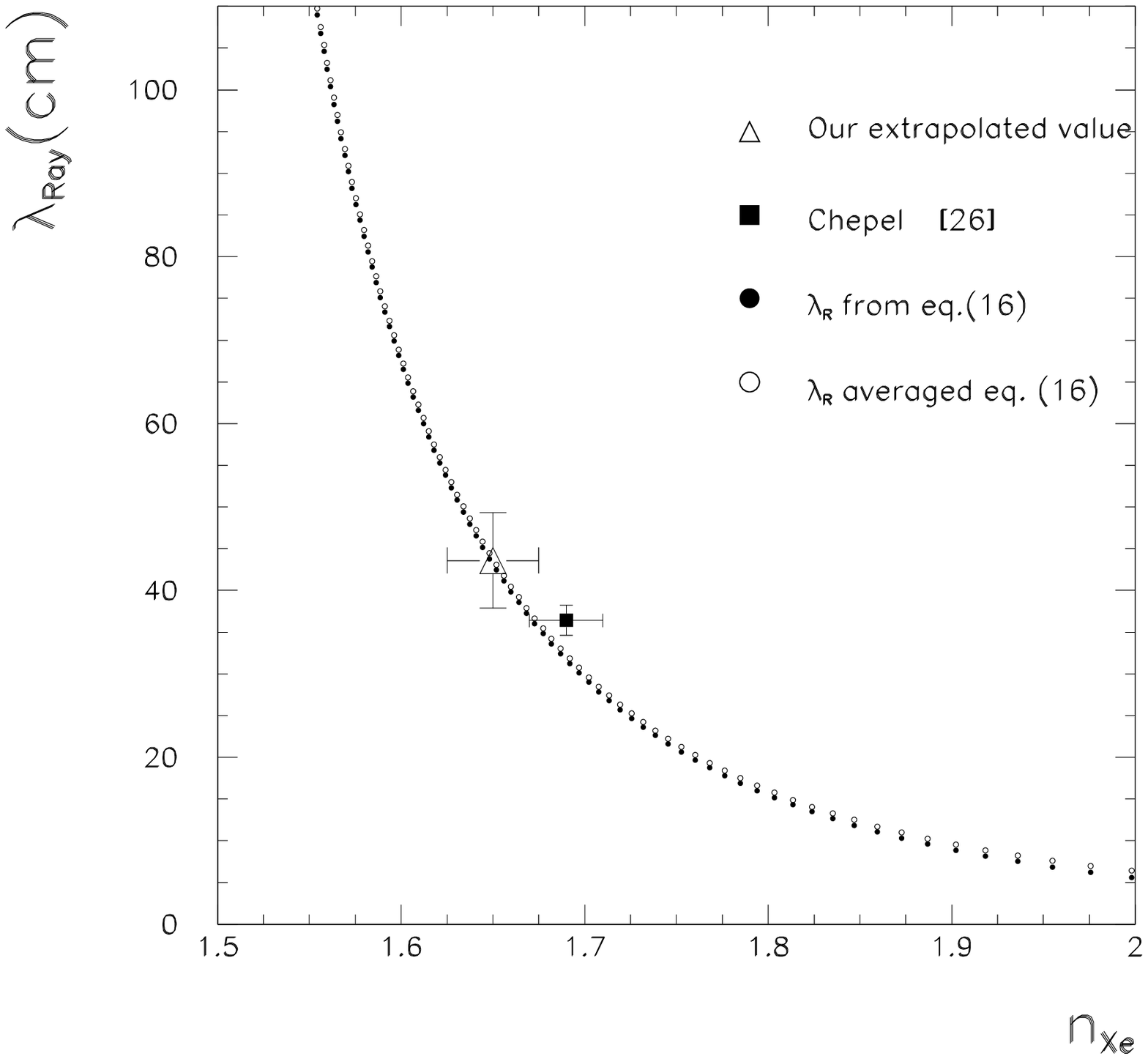}
	\end{center}
\vspace{-0.5cm}
\caption{$\lambda_R$ vs $n$ from equation \ref{eqn:h3} (black points). The average over the LXe emission spectrum is also shown (open points).
The only available experimental measurement of both $\lambda_R$ and $n$ is compared with the prediction of our model.}
\label{fig:pap15}
\end{figure}

$\lambda_R$ vs $n$ from equation (\ref{eqn:h3}) (black circles) is presented in fig.\ref{fig:pap15}. The input values for the refractive index $n(\omega)$ are the ones obtained from a fit to the low pressure VUV measurements \cite{bid}. The simple Clausius-Mossotti relation (\ref{eqn:F2}) (and its linearity as a function of $\rho_m$) is assumed to be valid. 
 There are only two published measurements of the Rayleigh scattering length in LXe at $178~\rm{nm}$ which gave $\lambda_{R}=29 \pm 2 ~\rm{cm}$ \cite{ishida} and $\lambda_{R}=36.4 \pm 1.8 ~\rm{cm}$ \cite{chepel} (in this last case the refractive index was also measured); we note that the two values are far apart.
These determinations were obtained for light produced by radioactive source excitation of LXe. This means that the measured $\lambda_R$ is an average over the 14~nm fwhm LXe emission spectrum. In checking the validity of equation (\ref{eqn:h3}) an average over the LXe emission spectrum must therefore be performed. In fig.~\ref{fig:pap15} equation (\ref{eqn:h3}) is also shown after averaging over the LXe emission spectrum (open circles); the differences between the two curves are small. In the same figure the only available experimental measurement of both $n$ and $\lambda_R$ \cite{chepel} (black square) is compared with the extrapolated values for the same two quantities (open triangle) obtained, as already explained, from the low-pressure VUV data and the Clausius-Mossotti relation.

In the case of small amounts (ppm) of other rare gases, like Ar or Kr, added to Xe, (\ref{eqn:h3}) preserves its validity and $\lambda_R$ does not vary appreciably. This is not true in the case of absorbing contaminants, as it was shown during the tests of the large prototype detector \cite{proposal}.

\section{Conclusions}
\label{sec:conclusions}
The experimental determination of LXe optical properties at the $\lambda=178$~nm emission line is not sufficiently complete and reliable. The literature on optical characteristics of Xe in other spectral regions and at various densities is critically examined. These data can be used to evaluate the refractive index $n$ and the Rayleigh scattering length $\lambda_R$ in the VUV for LXe since the validity, with good approximation, of the simple Clausius-Mossotti relation is confirmed from dilute gas to liquid densities.
A useful relation is derived which directly links $n$ and $\lambda_R$.

\section*{Acknowledgements}
Useful discussions with Prof. Giuseppe Grosso, Dept. of Physics, Pisa, are gratefully acknowledged.


\begin{thebibliography}{99}
\bibitem{proposal} MEG proposal to PSI, \ Proposal to INFN at
http://meg.web.psi.ch/docs/index.html.
\bibitem{mihara} S. Mihara {\it et al.}, \emph{IEEE TNS}, {\bf 49}, 588 (2002).
\bibitem{crc} {\it Handbook of Chemistry \& Physics}, The Chemical Rubber Company.
\bibitem{sin69} A.C. Sinnock and B.L. Smith, \emph{Phys. Rev.}, {\bf 181}, 1297 (1969).
\bibitem{airliquide} L'Air Liquide--Division Scientifique, {\it Encyclop\'edie 
des gaz}, Elsevier, Amsterdam (1976).
\bibitem{pdb} Particle Data Group: K. Hagiwara {\em et al.}, Phys. Rev. D66, 010001 (2002). 
\bibitem{bar} L.M. Barkov {\it et al.}, \emph{Nucl. Istr. and Meth.}, {\bf A379}, 482 (1996).
\bibitem{lan} G.M. Seidel, R.E. Lanou and W. Yao, \emph{Nucl. Instr. and Meth.}, {\bf A489}, 189 (2002).
\bibitem{doke:portugal} T.~Doke, Port.~Phys., {\bf 12}, 9 (1981), reprinted
in {\it Experimental Techniques in High Energy Physics}, T.~Ferbel ed.\@
(Addison Wesley, 1987).
\bibitem {doke} T. Doke and K. Masuda \emph{Nucl. Instr. and Meth.}, {\bf A420}, 62 (1999).
\bibitem{jortner:article}
J.~Jortner {\em et al.}, \emph{J.~Chem.~Phys.}, {\bf 42}, 4250 (1965).
\bibitem{jortner:book}
N.~Schwenter, E.E.~Koch and J.~Jortner, 
``Electronic Excitations in Condensed Rare Gases'',
Springer-Verlag, Berlin 1985.
\bibitem{ishida} N.~Ishida {\it et al.}, \emph{Nucl. Instr. and Meth.}, 
{\bf A384}, 380 (1997).
\bibitem{ncs} S.F.~Mughabghab, M.~Divadenam and N.E.~Holden, {\it Neutron 
Cross Sections}, Academic Press New York (1981).
\bibitem {ber} R. Bernabei {\it et al.}, \emph{Nucl. Instr. and Meth.}, {\bf A482}, 729 (2002).
\bibitem{hohm} U.~Hohm and K.~Kerl \emph{Molecular Physics}, {\bf 69}, 803 (1990).
\bibitem{sin} A.C. Sinnock, \emph{J. Phys. C}, {\bf 13}, 2375 (1980).
\bibitem{bid} A. Bideau-Mehu {\it et al.}, \emph{J. Quant. Spectrosc. Transfer}, {\bf 25}, 395 (1981).
\bibitem{laporte} P. Laporte and I.T. Steinberger, \emph{Phys. Rev. A}, {\bf 15}, 2538 (1977).
\bibitem{landau} L.D.~Landau and E.M.~Lifshitz, \emph{Electrodynamics of continuous media}, Pergamon Press.
\bibitem{jack} J.D. Jackson, \emph{Classical Electrodynamics}, Wiley.
\bibitem{huot} J. Hout and T.K. Bose, \emph{J. Chem. Phys.}, {\bf 95}, 2683 (1991).
\bibitem{acht}H.J. Achtermann {\it et al.}, \emph{J. Chem. Phys.}, {\bf 98}, 2308 (1993).
\bibitem{hohm1} U. Hohm, \emph{Molecular Physics}, {\bf 81}, 157 (1994).
\bibitem{subtil} J.L. Subtil {\it et al.}, \emph{Phys. Status Solidi}, {\bf B143}, 783 (1987).
\bibitem{chepel} V.~N.~Solovov, V.~Chepel, M.~I.~Lopes and A.~Hitachi,
Nucl.\ Instrum.\ Meth.\ A {\bf 516}, 462 (2004)
[arXiv:physics/0307044].
\bibitem{rice} S.A. Rice and J. Jortner, \emph{J.~Chem.~Phys.}, {\bf 44}, 4470 (1966).
\bibitem{steinberger} I.T. Steinberger and U. Asaf, \emph{Phys. Rev. B}, {\bf 8}, 914 (1973).
\end{thebibliography}
\end{document}